\documentclass[10pt]{article}
\usepackage[left=2cm,top=3cm,right=2cm,bottom=2cm]{geometry}
\usepackage[]{natbib}      
\usepackage{url}
\setcitestyle{semicolon,numbers,square}
\usepackage{dcolumn}     
\usepackage{multirow}    
\usepackage{rotating}    
\usepackage{graphicx}    
\usepackage{bm}          

\author{Robert W. Johnson \\
\small Alphawave Research\\[-0.8ex]
\small Atlanta, GA, USA\\
\small \texttt{robjohnson@alphawaveresearch.com}\\}
\title{Fitting a sum of exponentials to lattice correlation functions using a non-uniform prior}
\date{September 3, 2010\\ 
\small keywords: torelon spectrum, glueball spectrum, excited states\\
PACS: 11.15.Ha, 12.38.Lg, 12.39.Mk}

\newcommand{\beq}{\begin{equation}}
\newcommand{\eeq}{\end{equation}}
\newcommand{\bea}{\begin{eqnarray}}
\newcommand{\eea}{\end{eqnarray}}
\newcommand{\rmd}{{\mathrm d}}
\newcommand{\mcal}[1]{\mathcal{#1}}
\newcommand{\eref}[1]{(\ref{#1})}
\newcommand{\del}{\nabla}
\newcommand{\ber}{\begin{sideways}}
\newcommand{\eer}{\end{sideways}}
\newcommand{\dfrac}[2]{\displaystyle\frac{#1}{#2}}
\newcommand{\bra}[1]{\langle #1 \vert}
\newcommand{\ket}[1]{\vert #1 \rangle}
\newcommand{\tens}[1]{\mathbf{#1}}

\newcolumntype{d}[1]{D{.}{.}{#1}}

\begin{document}
\maketitle
\begin{abstract}
Excited states are extracted from lattice correlation functions using a non-uniform prior on the model parameters.  Models for both a single exponential and a sum of exponentials are considered, as well as an alternate model for the orthogonalization of the correlation functions.  Results from an analysis of torelon and glueball operators indicate the Bayesian methodology compares well with the usual interpretation of effective mass tables produced by a variational procedure.  Applications of the methodology are discussed.
\end{abstract}

\includegraphics[angle=30,width=6.5in]{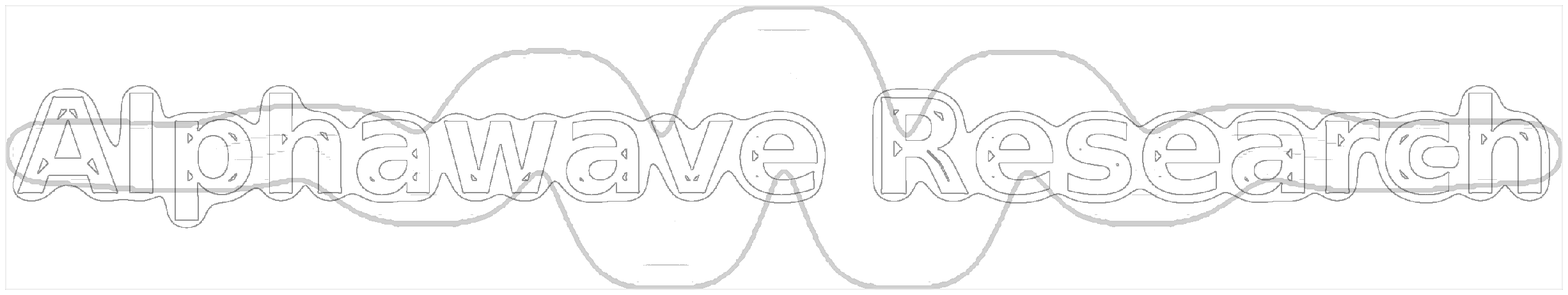}

\twocolumn


\section{Introduction}
\label{sec:intro}
The best means by which to extract the mass of a state from its lattice correlation function remains an open question.  Common practice still relies on visual examination of an effective mass table for evidence of a mass plateau to identify the mass spectrum of a particular class of operator.  A single correlation function will contain contributions from more than one mass eigenstate even after application of a variational procedure, as the lightest state will only dominate at long time intervals.  Further complications arise from the statistical noise of an actual simulation.   Here, we address the question of fitting a sum of exponentials to the correlation functions through the application of Bayesian data analysis with a non-uniform prior on the model parameters.

While fitting a sum of exponentials has now been superseded by the spectral maximum entropy method (MEM)~\citep{Asakawa:2001459,Yamazaki:014501,Langfeld:2002131,Kanaya:2003233c,Fiebig:094512}, it remains in use~\citep{Morningstar:2002185,expfitnmr2:2005} when one has reason to believe the data may be described by a discrete spectrum, as to be expected for a finite lattice~\citep{Montvay:1994cy}.  The use of least squares, or maximum likelihood analysis, to fit the model parameters contains an unrecognized bias on the magnitude of the mass.  The Bayesian methodology with a non-uniform prior allows one to correct for that bias as well as to encode additional relevant information.  First we will discuss our choice of prior for the amplitude and decay constant of an exponential fitting function.  Considering four models of exponential decay appropriate for lattice correlation functions, we apply the methodology to a collection of torelon operators, followed by a discussion of how one determines the most likely model.  We next consider an alternate approach, based upon modelling the orthogonalization provided by the variational procedure, and apply that method to both torelon and glueball correlation functions.  We close by summarizing our results for the spectra and discussing the utility of these algorithms.

The correlation functions used herein come primarily from a 10,000 measurement run of a lattice simulation at $\beta = 6$ and $L=16$ for SU(2) pure gauge theory in $D=2+1$ dimensions using the Wilson action with spacing $a$.  Masses are given in terms of lattice units throughout.  Evaluation of operators occured once every 10 compound sweeps after sufficient thermalization, updated via the Kennedy-Pendleton heat bath algorithm~\citep{Kennedy:1985nu} augmented with a 4:1 ratio of over-relaxation sweeps~\citep{Creutz:1987xi} and global gauge transformations.  Further simulation details and particulars of the superlink method of operator construction are found in Ref.~\citep{johnson:074502}.  Here we will consider operators for the torelon constructed from Polyakov loops and operators for the $J^P=0^+$ glueball constructed from square boxes of diagonal superlinks.  The methodology applies equally to general gauge groups in arbitrary dimension once the correlation functions are computed; this paper concerns itself not with the presentation of new results but rather with an investigation of a new technique for getting those results.

\section{Choice of prior}
\label{sec:prior}
The essential feature of Bayesian data analysis which takes it beyond simple least-squares fitting is the use of a non-uniform prior in appropriate circumstances~\citep{Sivia:1996}.  Using the language of conditional probabilities~\citep{Durrett:1994}, we write ``the probability of $A$ given $B$ under conditions $I$'' as \beq
{\rm prob}(A|B;I) \equiv p(A |_I B) \equiv p^A_B
\eeq when the background information $I$ is unchanging, and one states Bayes' theorem in the context of parameter estimation as \beq \label{eqn:Bayes}
p^{\vec{X}}_D = p^{\vec{X}} p^D_{\vec{X}} / p^D \;,
\eeq reading ``the evidence for parameters $\vec{X}$ given data $D$ equals the prior for $\vec{X}$ times the likelihood of data $D$ given $\vec{X}$ divided by the chance of measuring data $D$''.  What we call ``the evidence'' is often called ``the posterior'', as the normalization constant $p^D$ affecting neither parameter estimation nor model selection is sometimes called ``evidence''; both ``prior'' and ``likelihood'' have their usual meaning.  The logarithm (base $e$) of Eq.~\eref{eqn:Bayes} reads $L_E = L_P + L_L + \#_D$, where the final term is a constant equal to $- \log p^D$.  For independent data $\vec{D} = \{D_t\}$ indexed by $t$ with Gaussian noise $\vec{\sigma}$, the likelihood factors as $p^{\vec{D}}_{\vec{X}} = \prod_t (2 \pi \sigma_t^2)^{-1/2} \exp (-R_t^2/2)$, where $R_t \equiv [M_t(\vec{X}) - D_t]/\sigma_t$ is the normalized residual of the model $M$, so that $L_L$ has one term proportional to the measure of fit $\chi^2 = \sum_t R_t^2$ and another which is constant.  With the definition of the merit function in terms of the model parameters, \beq
-L^{\vec{X}} -L^{\vec{D}}_{\vec{X}} = - \sum_{x \in \vec{X}} \log p^x + \dfrac{1}{2} \chi^2 + \#_{\vec{\sigma}} \;,
\eeq the problem is reduced(!) to one of nonlinear global optimization, with all the attendant difficulties: just because a solution has not been found does not mean it cannot be found, and just because a (local) solution is found does not mean it is the global one.  Short of evaluating the merit function over the entire prior range, one must rely on intuition and luck to varying degrees.  One's intuition, encoded in the form and domain of the prior functions $p^{\vec{X}}$, contributes to the gradient of the log evidence in the limit of poor data, thereby improving the chances of success.

\begin{figure}
\includegraphics{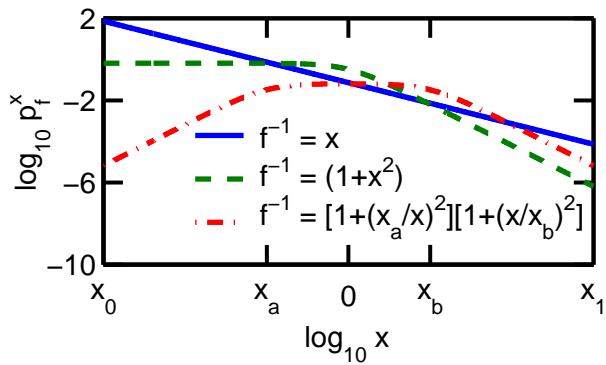}%
\caption{\label{fig:A} Comparison of normalized priors $p^x_f$ for various functions $f$}
\end{figure}

The choice of prior~\citep{dAgo:1998} represents one's background knowledge on the likely distribution of a parameter $x \in [x_0, x_1]$ before analysis of the current set of data, and a uniform prior $p^x_1 = \Delta^{-1}_x \equiv 1 / (x_1 - x_0)$ reduces Bayes' theorem to a statement of proportionality between the evidence and the likelihood, $p^x_D \propto p^D_x$.  A non-uniform prior $p^x_f$ arises naturally in many contexts, often representing a prior which is uniform over a change of variables $x \rightarrow F$ for some integrable function $f(x) = \rmd F / \rmd x$, with normalization $p^x_f = \Delta^{-1}_F f(x)$ for $\Delta_F \equiv \int_{x_0}^{x_1} f\; \rmd x$ such that $\int_{x_0}^{x_1} p^x_f\; \rmd x \equiv 1$.  Besides the uniform prior, one commonly encounters the Jeffreys' prior $f^{-1} = x$ uniform over $\log x$ and the Cauchy distribution $f^{-1} = 1 + x^2$ uniform over $\arctan x$, and we will find it useful to consider a prior we call the double Cauchy prior, $f^{-1} = [1+(x_a/x)^2][1+(x/x_b)^2]$ for $x_0 < x_a < x_b < x_1$, that mirrors the form of the Cauchy prior around a central region.  The scale parameters $x_a$ and $x_b$ allow one to introduce ``soft'' limits on the parameter $x$ well within the ``hard'' limits imposed by one's evaluation range and make explicit the choice of units for $x$.  These priors are compared in Fig.~\ref{fig:A}.

\begin{figure}
\includegraphics{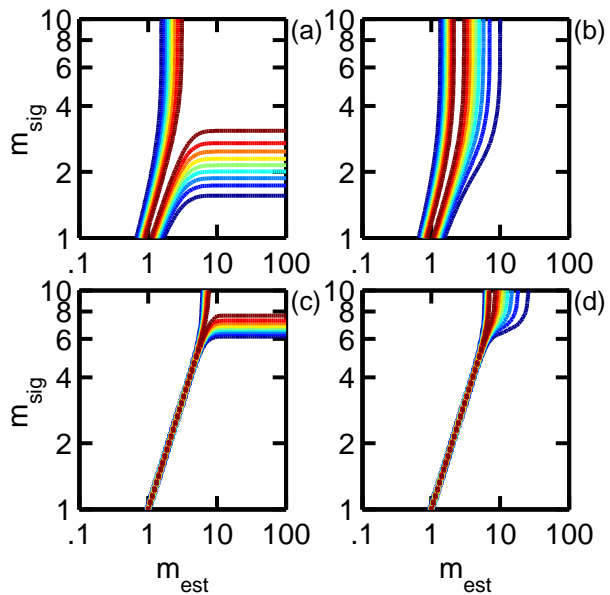}%
\caption{\label{fig:B} Contours of the evidence for the uniform prior (left) and the Cauchy prior (right) as the assumed variance decreases from $\sigma_t = 10^{-1}$ in (a) and (b) to $\sigma_t = 10^{-3}$ in (c) and (d).  The known mass of the test signal is $m_\mathrm{sig}$ ranging from 1 to 10, and the estimate from the evidence is $m_\mathrm{est}$.  The uniform prior does not resolve the mass beyond a limit imposed by the quality of the data, while the Cauchy prior may underestimate the mass for poor data}
\end{figure}

When fitting a single exponential $y(t) = A \exp (- m t)$ to data, one commonly takes the logarithm of the ordinate to yield a linear model $L_y(t) = L_A - m t$ with parameters $L_A$ and $m$.  Identifying the intercept as a location parameter with uniform prior indicates a Jeffreys' prior for the amplitude $A$, and the prior uniform over the angle $\tan \theta = m$ is the Cauchy distribution.  Upon normalization $L_y(0) = 0$, only the slope $m$ remains, whose best estimate $m_\mathrm{est}$ from a noisy exponential with known decay $m_\mathrm{sig}$ is found by minimizing the merit function $-L_E = \chi^2/2 + \log (1+m^2) + \#$.  Maximum likelihood implies using a prior uniform on the magnitude of the slope $m$ which appears in $\theta$ as $p^\theta \propto 1 + (\tan \theta)^2$, clearly displaying a preference for a slope (or mass) of extreme magnitude.  In Fig.~\ref{fig:B} we compare the estimate using both the uniform (a) and the Cauchy prior (b) by displaying contours of the evidence, with assumed variance $\sigma_t = 10^{-1}$, which is not the same as the nonlinear noise added to make the pure exponential resemble an actual lattice correlation function, and with $\sigma_t = 10^{-3}$ in (c) and (d).  We see that the effect of the non-uniform prior $\del L_P \neq 0$ is to reduce the spread of the evidence beyond a value of $m_\mathrm{est}$ determined by the precision of the data, which in practice contributes to the gradient of the log evidence $\del L_E$ when the data has very little to say, $\del L_L \rightarrow 0$.

\begin{figure}
\includegraphics{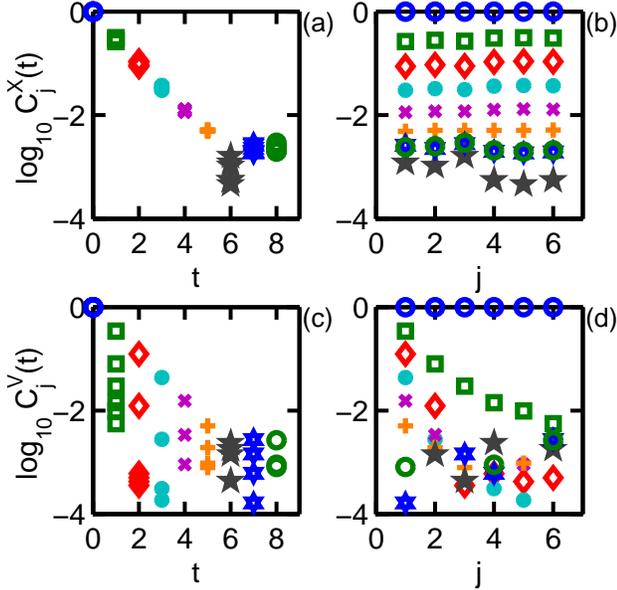}%
\caption{\label{fig:C} Positive values of the normalized self-correlation functions for torelon operators displayed against time $t$ and index $j$, where the symbol type indicates the timeslice separation.  The $C^V_j(t)$ in (c) and (d) have been orthogonalized by a variational procedure, while $C^X_j(t)$ in (a) and (b) are the correlations of the original basis.  The ground state is observed from the regularity in spacing for $j=1$ in (d)}
\end{figure}

\section{Analysis of torelon operators}
\label{sec:torelon}
Turning now to some real data, in Fig.~\ref{fig:C} we display on a logarithmic axis the values of the normalized correlation functions $C_j(t)$ indexed by $t$ in units of $a$ and averaged over both spatial directions of the timeslice for torelon operators constructed from smeared Polyakov loops non-contractible around the spatial lattice, noting that they do not tend to zero at the largest time separation.  Our count of operators equals 6 indexed by $j$, representing 3 step sizes and 2 smearing levels.  We denote by $C^X$ the self-correlations of the original basis of operators and  by $C^V$ the correlation functions produced by a variational procedure to enhance their orthogonality~\citep{Berg:1983109,luscher:1990222,teper-1999-59}.  Specifically, we take the modal matrix $\tens{E}$ of real eigenvectors found by diagonalizing the product of the first two cross-correlation matrices $\tens{X}_t \equiv X_{ij}(t) = \langle \mcal{O}_i^*(t+\tau) \mcal{O}_j(\tau) \rangle_\tau$ for zero-momentum timeslice operators $\mcal{O}(t)$ such that \beq \label{eqn:varieig}
\tens{X}^{-1}_0 \tens{X}_1 \tens{E} = \tens{E} \tens{D} \;,
\eeq where $\tens{D}$ is a diagonal matrix arranged by increasing eigenvalue.  Symmetrizing the cross-correlations on $i,j$ is equivalent to averaging them over both temporal orientations $\pm t$, so that $\tens{X}_t + \tens{X}_t^T = \tens{X}_t + \tens{X}_{L-t}$, and we will work with the symmetrized data with temporal range $t \in [0, L/2]$.  The orthogonal basis $\tens{O} \tens{E}$ is formed from the original operators $\tens{O} \equiv O_{tj} = \mcal{O}_j(t)$ so that $\tens{V}_t = \tens{E}^T \tens{X}_t \tens{E}$, and the variational self-correlators are extracted $C^V_j(t) = \delta_{ij} V_{ij}(t)$.  

\begin{table}
\caption{Effective mass table for original $a m^X_j$ and variational $a m^V_j$ torelon operators}
\label{tab:B} 
\addtolength{\tabcolsep}{-1pt}
\begin{tabular}{c|cccccccc}
\hline\noalign{\smallskip}
t &        1 &        2 &         3 &          4 &           5 &           6 &          7 &            8 \\
\noalign{\smallskip}\hline\noalign{\smallskip}
\multirow{6}{*}{\ber $a m^X_j(t)$ \eer} 
& 1.34 & 1.10 & 1.06 & 0.98 & 0.85 & 1.40 & 0.00 & 0.09 \\
& 1.28 & 1.09 & 1.06 & 1.01 & 0.84 & 1.58 & 0.00 & 0.00 \\
& 1.33 & 1.10 & 1.05 & 0.96 & 0.84 & 1.15 & 0.00 & 0.00 \\
& 1.18 & 1.06 & 1.07 & 1.05 & 0.92 & 2.23 & 0.00 & 0.00 \\
& 1.16 & 1.05 & 1.07 & 1.05 & 0.94 & 2.41 & 0.00 & 0.00 \\
& 1.18 & 1.06 & 1.07 & 1.04 & 0.92 & 2.24 & 0.00 & 0.00 \\
\noalign{\smallskip}\hline\noalign{\smallskip}
\multirow{6}{*}{\ber $a m^V_j(t)$ \eer} 
& 1.07 & 1.01 & 1.05 & 1.03 & 1.12 & 0.00 & 0.00 & 0.00 \\
& 2.53 & 1.87 & 1.49 & 0.00 & 0.56 & 0.31 & 0.00 & 0.00 \\
& 3.52 & 4.43 & 0.00 & 2.11 & 0.00 & 0.60 & 0.00 & 0.00 \\
& 4.26 & 3.16 & 0.67 & 0.00 & 0.00 & 0.00 & 1.34 & 0.00 \\
& 4.63 & 3.14 & 0.83 & 0.00 & 0.00 & 0.00 & 2.36 & 0.00 \\
& 5.19 & 2.41 & 0.00 & 2.66 & 0.00 & 0.00 & 0.00 & 0.03 \\
\noalign{\smallskip}\hline
\end{tabular}
\end{table}

Common practice is to form the effective mass table from the correlation functions, defined by $a m_j(t) \equiv \log [C_j(t-1)/C_j(t)]$ for $t \geq 1$, with errors given by jackknife analysis.  In Table~\ref{tab:B} we display the effective mass table for both the original and variational torelon operators, with negative and imaginary values zeroed out and neglecting the error analysis.  Clear evidence for a mass plateau at the ground state identified by the variational procedure is seen for the correlation functions of the original basis, but the remainder of the data is hard to interpret---how should one identify and extract the mass of the excited states?  In other words, how is the information obtained for the remaining states (which here appear to be multi-torelon excitations) by the variational procedure?  To address these questions, we consider fitting a sum of exponentials with free parameters for the amplitudes and decay constants, a notoriously hard problem~\citep{Acton:1970} whose difficulties we hope to mitigate through the use of non-uniform priors.

\subsection{Models for exponential decay}
The first of the models we consider, conveniently indexed by their number of parameters, is given by a single exponential with a free decay constant, \beq
M_1:\; C(t) = \exp (-m_1 t) \;,
\eeq which is driven primarily by the first non-constrained value $C(1)$.  As an alternative, we consider a model which utilizes a constant to represent the statistical noise of the simulation, \beq
M_2:\; C(t) = A_1 \exp (-m_1 t) + (1-A_1) \;,
\eeq chosing a double Cauchy prior for $m_1$ in both models with hard limits of 0.1 and 6 and soft limits of 1 and 4 in lattice units of mass.  (With such a tight range, the shape of the double Cauchy prior approaches that of an offset Gaussian on logarithmic axes.)  For the amplitude $A_1$, the Jeffreys' prior is chosen over range $[0.8,1.1]$.  Detailed observation of the correlation functions indicates that fluctuations have become dominant several sites before the midpoint of the lattice is reached, so we restrict the fitting window to a range $t_\mathrm{fit} \in [0,5]$ and neglect to replace the exponential with a hyperbolic cosine representing correlations the long way around a finite lattice, justified when $m L \gg 1$.  (The variance of the correlators at $t=0$ is calculated before the normalization is applied.)  In Table~\ref{tab:C} we give the results for these two single exponential models applied to the variational basis correlations $C^V_j(t)$ with the standard error indicated in parentheses and $\del_{10} L_E \equiv \log_{10} \vert \del L_E \vert$.

\begin{table}
\caption{Single exponential model parameters for variational basis torelon operators $C^V_j(t)$}
\label{tab:C} 
\begin{tabular}{d{6}d{2}|d{6}d{6}d{2}}
\hline\noalign{\smallskip}
\multicolumn{2}{c|}{$M_1$} & \multicolumn{3}{c}{$M_2$} \\
\noalign{\smallskip}\hline\noalign{\smallskip}
\multicolumn{1}{c}{$a m_1^V$} & \multicolumn{1}{c|}{$\del_{10} L_E$} & \multicolumn{1}{c}{$A_1$} & \multicolumn{1}{c}{$a m_1^V$} & \multicolumn{1}{c}{$\del_{10} L_E$} \\
\noalign{\smallskip}\hline\noalign{\smallskip}
1.071(4)   & -11.2 & 0.997(2) & 1.077(6)   & -11.1 \\
2.534(22)  &  -9.9 & 0.997(1) & 2.572(26)  & -10.7 \\
3.524(60)  & -14.3 & 1.000(1) & 3.514(67)  &  -9.9 \\
4.257(125) & -10.7 & 1.000(1) & 4.236(137) &  -9.8 \\
4.625(180) &  -7.6 & 0.999(1) & 4.684(212) & -10.3 \\
5.172(310) &  -8.3 & 1.001(1) & 5.058(312) & -11.0 \\
 \noalign{\smallskip}\hline
\end{tabular}
\end{table}

\begin{table*}
\caption{Double exponential model parameters for original basis torelon operators $C^X_j(t)$}
\label{tab:D} 
\addtolength{\tabcolsep}{-1pt}
\begin{tabular}{d{8}d{7}d{9}d{1}|d{8}d{8}d{7}d{9}d{1}}
\hline\noalign{\smallskip}
\multicolumn{4}{c|}{$M_3$} & \multicolumn{5}{c}{$M_4$} \\
\noalign{\smallskip}\hline\noalign{\smallskip}
\multicolumn{1}{c}{$A_1$} & \multicolumn{1}{c}{$a m_1^X$} & \multicolumn{1}{c}{$a m_2^X$} & \multicolumn{1}{c|}{$\del_{10} L_E$} & \multicolumn{1}{c}{$A_1$} & \multicolumn{1}{c}{$A_2$} & \multicolumn{1}{c}{$a m_1^X$} & \multicolumn{1}{c}{$a m_2^X$} & \multicolumn{1}{c}{$\del_{10} L_E$} \\ \hline
0.654(137) & 1.020(81) & 2.580(769)  & -11.5 & 0.751(87)  & 0.248(88)  & 1.089(63) & 3.487(1670) & -10.5 \\
0.732(128) & 1.033(70) & 2.655(993)  & -10.9 & 0.804(65)  & 0.195(66)  & 1.082(47) & 3.598(1782) &  -9.9 \\
0.614(162) & 0.993(95) & 2.316(600)  & -11.0 & 0.754(104) & 0.244(105) & 1.087(71) & 3.257(1579) & -10.9 \\
0.851(61)  & 1.042(34) & 3.242(1653) & -11.7 & 0.869(37)  & 0.131(38)  & 1.054(30) & 3.846(1918) & -10.1 \\
0.868(55)  & 1.039(31) & 3.268(1714) & -11.5 & 0.883(34)  & 0.117(35)  & 1.047(28) & 3.840(1929) & -10.8 \\
0.855(63)  & 1.041(35) & 3.182(1644) & -12.0 & 0.874(37)  & 0.125(38)  & 1.053(29) & 3.824(1919) & -10.5 \\
\noalign{\smallskip}\hline
\end{tabular}
\end{table*}

To accommodate the covariance in $t$ of the functions $C_j(t)$, one generalizes the measure of fit by \beq \label{eqn:genchi}
\chi^2_j = (\vec{M}_j-\vec{C}_j)^T \tens{W}_j (\vec{M}_j-\vec{C}_j) \;,
\eeq where $\vec{M}_j$ and $\vec{C}_j$ are column vectors indexed by $t$ and $\tens{W} \equiv \tens{\Sigma}^{-1}$ is the inverse of the variance matrix evaluated~\citep{Montvay:1994cy} from the $N_n$ configurations indexed by $n$, \beq
\tens{\Sigma} \equiv \Sigma_{tu} = (N_n-1)^{-1} \bigl \langle (C^n_t - \langle C^n_t \rangle_n)(C^n_u - \langle C^n_u \rangle_n) \bigr \rangle_n \;.
\eeq  Note that it is the \emph{matrix} inverse~\citep{Sivia:1996} which appears in Eqn.~(\ref{eqn:genchi}), a distinction lost in the more common notation of the general measure of fit as a sum over indices.

\begin{figure}
\includegraphics{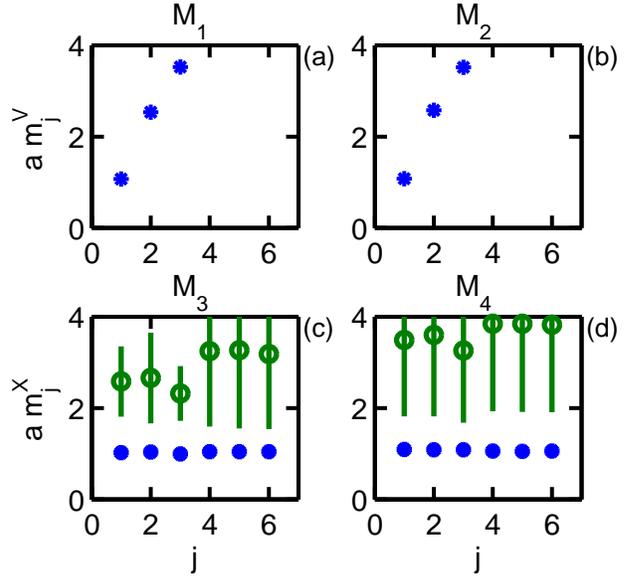}%
\caption{\label{fig:D} Comparison of the spectra of the torelon operators indexed by state $j$ for the various models}
\end{figure}

Considering now a sum of two exponentials, the third model is just the simple sum without a noise floor, \beq
M_3:\; C(t) = A_1 \exp (-m_1 t) + (1-A_1) \exp (-m_2 t) \;,
\eeq and the fourth has two free amplitudes with a noise floor, \beq
M_4:\; C(t) = A_1 \exp (-m_1 t) + A_2 \exp (-m_2 t) + (1-A_1-A_2) \;.
\eeq  We maintain the double Cauchy prior on the masses and the Jeffreys' prior on the amplitudes, this time with range $[0.1,1]$.  These models are applied to the original basis of correlation functions $C^X_j(t)$.  Their solution parameters are displayed in Table~\ref{tab:D}.  We see that the original self-correlators do contain evidence of at least one excitation; however, the resolution is poor in terms of the standard error.  When compared graphically in Fig.~\ref{fig:D}, where for the sum of exponentials models the secondary states are marked by open circles while the dominant states are filled circles, we find a consistent estimate for the ground state, whereas the estimate for the excited states depends upon the use of the noise floor constant when evaluated from $C^X_j(t)$.

\subsection{Model selection}
\label{sec:compare}
How does one compare the quality of fit between the models?  The Bayesian formalism addresses model selection by considering the ratio of the evidence for each model with no prior preference $p^A = p^B$, thus reducing to the likelihood ratio \beq \label{eqn:evirat}
\dfrac{p^{A}_D}{p^{B}_D} = \dfrac{p^{A} p^D_{A} / p^D}{p^{B} p^D_{B} / p^D} \rightarrow \dfrac{p^D_A}{p^D_B} \;,
\eeq whose factor for each model is the ``integrated probability bump'' over the model parameters $\vec{X}$, \beq \label{eqn:modellike}
p^D_{M} = \int p^{D,\vec{X}}_{M} \rmd \vec{X} = \int p^{\vec{X}}_{M} p^D_{\vec{X},M} \rmd \vec{X} \;,
\eeq (normalized by the chance of $D$ which factors out of the ratio).  Our choice of nomenclature [\emph{cf}. Eqns.~(\ref{eqn:Bayes}) and (\ref{eqn:evirat})] identifies Eqn.~(\ref{eqn:modellike}) as ``the likelihood of $D$ given $M$'' (not ``the chance of $D$ given $M$'') which marginalizes into a product of the prior for $\vec{X}$ and the likelihood for $\vec{X}$.  In other words, the evidence for the model is the (unnormalized) integral of the evidence for its parameters.  Introducing a model subscript $M$ to Eqn.~(\ref{eqn:Bayes}), one sees that $p^D_M$ is the normalizing constant such that $\int p^{\vec{X}}_{D, M} d\vec{X}$ equals unity.  An unfortunate confusion of nomenclature arises because $p^D_M$ appears both in the position of chance in Eqn.~(\ref{eqn:Bayes}) and in the position of likelihood in Eqn.~(\ref{eqn:evirat}).

Under the quadratic approximation, generally acceptable when the evidence is not severely truncated by the prior range, one can evaluate the integral analytically to write the negative logarithm of the likelihood as \beq
-L^D_M \approx \dfrac{1}{2} \chi^2 + \sum_k \log f^{-1}_k - \sum_k \log \left( \Delta^{-1}_k \sqrt{2 \pi / h_k} \right) \;,
\eeq for $\vec{X}$ indexed by $k$ and $\{h_k\}$ the eigenvalues of the inverse variance $\prod_k h_k = \mathrm{det}\, \tens{\Sigma}_{\vec{X}}^{-1}$, where the first two terms are the value of the merit function evaluated at its minimum and the remainder comprise the Occam factor accounting for the ratio of the width of the evidence $\tens{\Sigma}_{\vec{X}}$ to the prior range $\{\Delta_k\}$.  An additional parameter must provide not just a better fit but a significantly better fit in order for its plausibility to increase.  With several models to choose from, the one with the lowest value of $-L^D_M$ is deemed the most plausible, with the relative probability given by the exponential of the difference between the (negative) log evidence for each.  (Some investigators take the further step of normalizing $\sum_M p^D_M = 1$ which we neglect.)  In Table~\ref{tab:E} we display $\{-L^D_M\}_j$ for our four models, noting that, except for one case, the models $M_1$ and $M_3$ without the noise constant are preferred over those with its inclusion.

\begin{table}
\caption{Negative log evidence of the models for torelon operators}
\label{tab:E} 
\begin{tabular}{c|d{1}d{1}|d{1}d{1}}
\hline\noalign{\smallskip}
\multicolumn{1}{r|}{model} & \multicolumn{1}{c}{$M_1$} & \multicolumn{1}{c|}{$M_2$} & \multicolumn{1}{c}{$M_3$} & \multicolumn{1}{c}{$M_4$} \\
\noalign{\smallskip}\hline\noalign{\smallskip}
\multirow{6}{*}{\ber $-L^D_M$ \eer} 
& 16.5 & 19.2 & 10.3 & 14.4 \\
& 12.1 & 11.3 & 10.3 & 14.4 \\
&  4.0 &  8.8 & 10.3 & 14.4 \\
&  3.3 &  8.2 & 10.4 & 14.4 \\
&  3.0 &  7.7 & 10.5 & 14.4 \\
&  3.0 &  7.8 & 10.4 & 14.4 \\
\noalign{\smallskip}\hline
\end{tabular}
\end{table}

\begin{figure}
\includegraphics{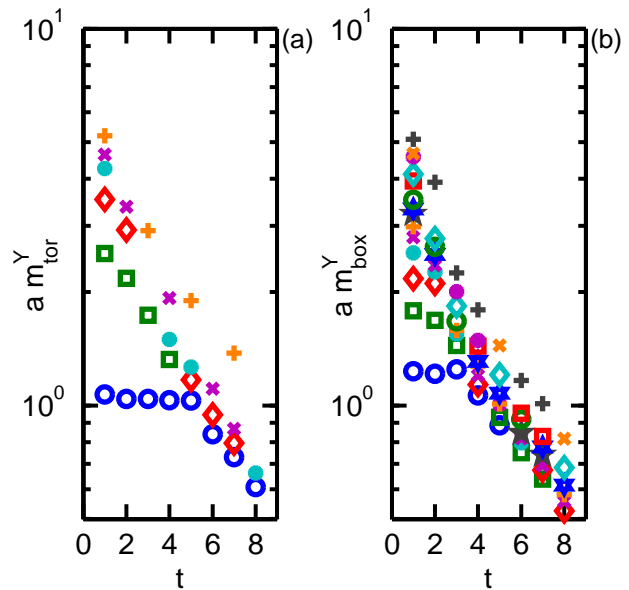}%
\caption{\label{fig:E} Effective masses from singular value decomposition for torelon (a) and $0^+$ glueball operators (b) indexed by time-slice separation $t$, where only positive eigenvalues are marked.  The symbol type indicates the ordering of the states $j$, and the estimate for the ground state $\circ$ is observed from the lowest horizontal values}
\end{figure}

\section{A better approach}
\label{sec:compofit}
The methodology of the previous section, while producing a credible analysis of the torelon data, did not fare well when faced with glueball correlation functions.  Consequently, we have investigated an alternate approach which takes into account the orthogonality of the mass eigenstates.  The focus of our investigation is a comparison of the two means by which one may estimate parameters from a set of data, either by inverting the data or by inference from the data~\citep{Sivia:1996,Press:1992}.  There are a variety of ways to invert a set of correlation functions to produce effective masses~\citep{Fiebig:094512}, but let us consider one based upon the variational method mentioned previously.  Rather than using the eigenvectors of Eqn.~(\ref{eqn:varieig}), one can apply the diagonalization in succession to the original basis of cross-correlators for $t > 0$, \beq \label{eqn:varimass}
\tens{X}^{-1}_0 \tens{X}_t \tens{E}_t = \tens{E}_t \tens{D}_t \;,
\eeq and then define effective masses in terms of the eigenvalues $a m_j^\mathrm{eff}(t) = - t^{-1} \log D_{jj}(t)$.  One finds that the same eigenvalues are produced by a two step procedure where one first orthogonalizes the basis according to the zero separation correlators, \beq \label{eqn:varidata}
\tens{Y}_t \equiv (\tens{A}_0 \tens{D}_0^{-1/2})^T \tens{X}_t \tens{A}_0 \tens{D}_0^{-1/2} \;,
\eeq where $\tens{X}_0 = \tens{A}_0 \tens{D}_0 \tens{A}_0^T$ so that $\tens{Y}_0 = \tens{I}$, and then inverts the data $\tens{Y}_t = \tens{E}_t \tens{D}_t \tens{E}_t^T$ to get the effective masses $a m_j^Y(t) = - t^{-1} \log D_{jj}(t)$.  As $\tens{Y}_t$ is symmetric, one may as well use singular value decomposition $\tens{Y}_t = \tens{A} \tens{S} \tens{B}^T$, where the columns of $\tens{A}$ and $\tens{B}$ may differ only by a sign for negative eigenvalues $D_{jj} = - S_{jj}$, to keep the values ordered by magnitude for display.  We see in Fig.~\ref{fig:E} how the mass eigenstates peel away at small $t$ from the band of values we interpret as simulation noise.  The essential difference between Eqns.~(\ref{eqn:varimass}) and (\ref{eqn:varidata}) is that the orthogonalization of the original basis $\tens{X}_t$ is done asymmetrically in (\ref{eqn:varimass}), coming in from the left in the guise of $\tens{X}^{-1}_0$, whereas in (\ref{eqn:varidata}) all $\tens{Y}_t$ are symmetric.  It is from data $\tens{Y}_t$ that we shall infer values for the mass and amplitude parameters.

\subsection{Modelling the variational procedure}
To model the orthogonalization provided by the variational procedure, we recognize that the step $\tens{X}_t \rightarrow \tens{Y}_t$ produces an orthogonal basis of operators which are not necessarily mass eigenstates.  The estimates for the mass and amplitude parameters should have no temporal dependence, and so we write the model as \beq
\tens{Y}_t \equiv \tens{A} \tens{D}_t \tens{A}^T \;,
\eeq where $\tens{A}^T \tens{A} = \tens{I}$ and $\tens{D}_t$ is the decay matrix $D_{jj}(t) = \exp(-t m_j)$.  In principle, the model can handle extracting a subset of eigenstates with the inclusion of some identity matrices, \beq
\tens{Y}_t \equiv \tens{A} ( \tens{D}_t - \tens{I}_k ) \tens{A}^T + \tens{I}_d
\eeq for $k < d$ and $\tens{A}$ of $d$ rows and $k$ columns, but we did not have much success with its convergence and will focus on the case $k = d$ for a subset of the entire data, $Y_{ij}(t)$ for $i, j \leq d$, ordered by decreasing value of their $t=1$ self-correlator.

While there are many options for representing the orthogonal amplitude matrix $\tens{A}$, we choose to use the minimal number of degrees of freedom provided by the composite parametrization~\citep{Spengler-290410} presented by Spengler \emph{et al}.  Summarized for the real case here, one starts with a vector $\vec{\Theta}$ of $d (d-1) / 2$ angles with domain $\Theta_l \in [-\pi,\pi)$ mapped into $\tens{\Theta} \equiv \Theta_{mn}$ using $l = n - d + m (2 d - m - 1) / 2$ for $1 \leq m < n \leq d$.  Using bra-ket matrix notation, the generators are defined by $\tens{\Lambda}_{mn} \equiv -i \ket{m} \bra{n} + i \ket{n} \bra{m}$.  The amplitude matrix is then given by the ordered product \beq
\tens{A} = \prod_{m=1}^{d-1}\left[ \prod_{n=m+1}^d \exp(i \Theta_{mn} \tens{\Lambda}_{mn}) \right] \;,
\eeq and a uniform prior is selected for the angles $\Theta_{mn}$.  In order to keep the mass estimates distinct, we parametrize their energy gaps above the vacuum $E_0 \equiv 0$ such that $m_j = \sum_{i \leq j} E_i$ for $j>0$.  In terms of the energy gaps, the decay matrix reads $\tens{D}_t = \sum_j \exp(-t \sum_i^j E_i) \ket{j} \bra{j}$.  The double Cauchy prior is chosen for the gap parameters, the first with hard limits of 0.1 and 4 and soft limits of 0.5 and 1 (1.5 for the glueball operators), and the rest with limit pairs of $[0.01,2]$ and $[0.1,1]$.

\subsection{Fitting the data}
This parametrization of the orthogonalized cross-correlators is unique only up to some leftover real phases, $\tens{Y}_t = \tens{A} \tens{D}_t \tens{A}^T = \tens{B} \tens{D}_t \tens{B}^T$ where $\tens{B} = \tens{A} \tilde{\tens{I}}$ for $\tilde{\tens{I}} = \sum_k \pm \ket{k} \bra{k}$.  Our optimizer did not like the restriction of the parameter domain to $[0,\pi)$, so we address the redundancy of the model by including a multi-modal factor of $2^d$ on the integrated probability bump.  The measure of fit is now a matrix quantity, \beq
\chi^2_{ij} = (\vec{Y}^M_{ij}-\vec{Y}^D_{ij})^T \tens{W}_{ij} (\vec{Y}^M_{ij}-\vec{Y}^D_{ij}) \;,
\eeq which we reduce to a scalar using the 1-norm \beq
\chi^2 \equiv \Vert \chi^2_{ij} \Vert_1 = \sum_{i=1}^d \sum_{j=1}^d \chi^2_{ij} \;.
\eeq  This formula is identical to that for the variance of a sum of quantities with covariance~\citep{Durrett:1994} which we will need to give error bars to our estimates for the eigenstate masses $m_j$.

\begin{table*}
\caption{Best fit parameters for torelon and $0^+$ glueball operators with the standard error below the parameter value}
\label{tab:F} 
\addtolength{\tabcolsep}{-1pt}
\begin{tabular}{c|c|d{3}d{3}d{3}d{3}|d{3}d{3}d{3}d{3}d{3}d{3}|d{1}d{1}}
\hline\noalign{\smallskip}
\multicolumn{1}{c|}{$N_n$} & \multicolumn{1}{c|}{$\mathcal{O}$} & \multicolumn{1}{c}{$E_1$} & \multicolumn{1}{c}{$E_2$} & \multicolumn{1}{c}{$E_3$} & \multicolumn{1}{c|}{$E_4$} & \multicolumn{1}{c}{$\Theta_1$} & \multicolumn{1}{c}{$\Theta_2$} & \multicolumn{1}{c}{$\Theta_3$} & \multicolumn{1}{c}{$\Theta_4$} & \multicolumn{1}{c}{$\Theta_5$} & \multicolumn{1}{c|}{$\Theta_6$} & \multicolumn{1}{c}{$\del_{10} L_E$} & \multicolumn{1}{c}{$- L_E$} \\
\noalign{\smallskip}\hline\noalign{\smallskip}
\multirow{4}{*}{\ber 10k \eer}
 & tor      &  1.083 &  1.440 &  1.659 &  0.509 & -0.347 &  0.149 &  0.177 &  0.508 &  0.480 & -0.871 & -10.8 & 133.3 \\
 & $\sigma$ &  0.004 &  0.018 &  0.094 &  0.176 &  0.004 &  0.003 &  0.003 &  0.018 &  0.016 &  0.176 & & \\
\noalign{\smallskip}\cline{2-14}\noalign{\smallskip}
 & box      &  1.271 &  0.478 &  0.597 &  0.468 &  0.536 & -0.251 & -0.052 &  0.739 &  0.289 &  2.365 & -8.6 & 172.3 \\
 & $\sigma$ &  0.005 &  0.005 &  0.013 &  0.018 &  0.011 &  0.006 &  0.002 &  0.016 &  0.008 &  0.039 & & \\
\noalign{\smallskip}\hline\noalign{\smallskip}
\multirow{4}{*}{\ber 1k \eer}
 & tor      &  1.141 &  1.638 &  0.721 &  0.971 & -0.366 &  0.282 & -0.112 &  0.787 & -0.213 & -1.078 & -11.3 &  88.2 \\
 & $\sigma$ &  0.008 &  0.042 &  0.100 &  0.239 &  0.007 &  0.006 &  0.005 &  0.051 &  0.051 &  0.103 & & \\
\noalign{\smallskip}\cline{2-14}\noalign{\smallskip}
 & box      &  1.278 &  0.563 &  0.599 &  0.403 &  0.564 & -0.211 & -0.062 &  0.830 &  0.358 & -0.562 & -11.6 & 101.3 \\
 & $\sigma$ &  0.013 &  0.024 &  0.042 &  0.071 &  0.019 &  0.016 &  0.011 &  0.039 &  0.028 &  0.094 & & \\
 \noalign{\smallskip}\hline
\end{tabular}
\end{table*}

\begin{table*}
\caption{Mass and amplitude parameters for torelon and $0^+$ glueball operators}
\label{tab:G} 
\begin{tabular}{c|c|d{3}d{3}d{3}d{3}|d{3}d{3}d{3}d{3}}
\hline\noalign{\smallskip}
\multicolumn{1}{c|}{$\mathcal{O}$} 
 & \multicolumn{1}{c|}{$N_n$} & \multicolumn{4}{c|}{10k} & \multicolumn{4}{c}{1k} \\
\noalign{\smallskip}\hline\noalign{\smallskip}
 & \multicolumn{1}{c|}{$j$} & \multicolumn{1}{c}{1} & \multicolumn{1}{c}{2} & \multicolumn{1}{c}{3} & \multicolumn{1}{c|}{4} & \multicolumn{1}{c}{1} & \multicolumn{1}{c}{2} & \multicolumn{1}{c}{3} & \multicolumn{1}{c}{4} \\
\noalign{\smallskip}\cline{2-10}\noalign{\smallskip}
\multirow{6}{*}{\ber tor \eer}
 & $m_j$ &  1.083 &  2.523 &  4.182 &  4.691 &  1.141 &  2.779 &  3.500 &  4.471 \\
 & $\sigma$ &  0.004 &  0.018 &  0.096 &  0.165 &  0.008 &  0.042 &  0.095 &  0.234 \\
 \noalign{\smallskip}\cline{2-10}\noalign{\smallskip}
 & \multirow{4}{*}{$\tens{A}$} &  0.915 & -0.400 & -0.046 &  0.018 &  0.891 & -0.448 & -0.038 &  0.059 \\
 & &  0.331 &  0.679 &  0.645 & -0.113 &  0.342 &  0.568 &  0.203 & -0.721 \\
 & & -0.147 & -0.415 &  0.368 & -0.819 & -0.277 & -0.658 &  0.474 & -0.515 \\
 & & -0.176 & -0.455 &  0.668 &  0.562 &  0.112 &  0.210 &  0.856 &  0.459 \\
\noalign{\smallskip}\hline\noalign{\smallskip}
\multirow{6}{*}{\ber box \eer}
 & $m_j$ &  1.271 &  1.749 &  2.347 &  2.815 &  1.278 &  1.841 &  2.440 &  2.843 \\
 & $\sigma$ &  0.005 &  0.007 &  0.015 &  0.027 &  0.013 &  0.022 &  0.040 &  0.066 \\
 \noalign{\smallskip}\cline{2-10}\noalign{\smallskip}
 & \multirow{4}{*}{$\tens{A}$} &  0.832 &  0.512 & -0.208 &  0.054 &  0.825 &  0.478 &  0.299 & -0.042 \\
 & & -0.494 &  0.520 & -0.607 &  0.342 & -0.522 &  0.445 &  0.699 & -0.202 \\
 & &  0.248 & -0.622 & -0.372 &  0.643 &  0.209 & -0.671 &  0.417 & -0.576 \\
 & &  0.052 & -0.285 & -0.671 & -0.683 &  0.062 & -0.350 &  0.498 &  0.791 \\
 \noalign{\smallskip}\hline
\end{tabular}
\end{table*}

\begin{figure*}
\includegraphics{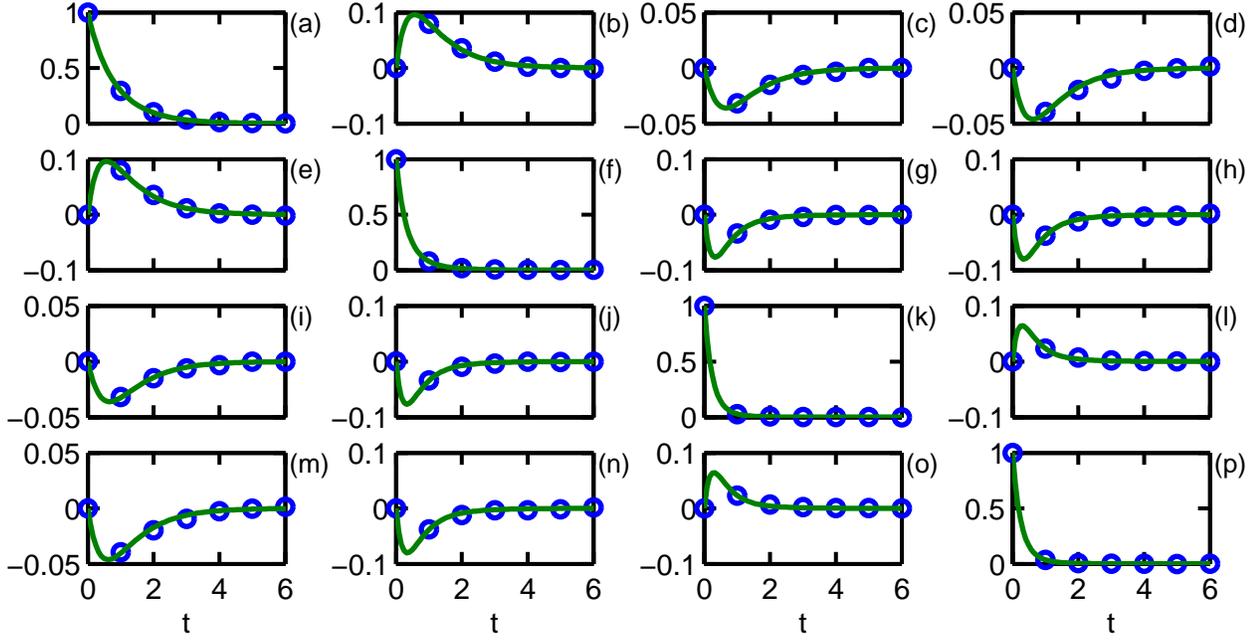}%
\caption{\label{fig:F} Comparison of $\tens{Y}_t \equiv Y_{ij}(t)$ for the model (solid) and data (circled) for the torelon operators using 10k data }
\end{figure*}

\begin{figure*}
\includegraphics{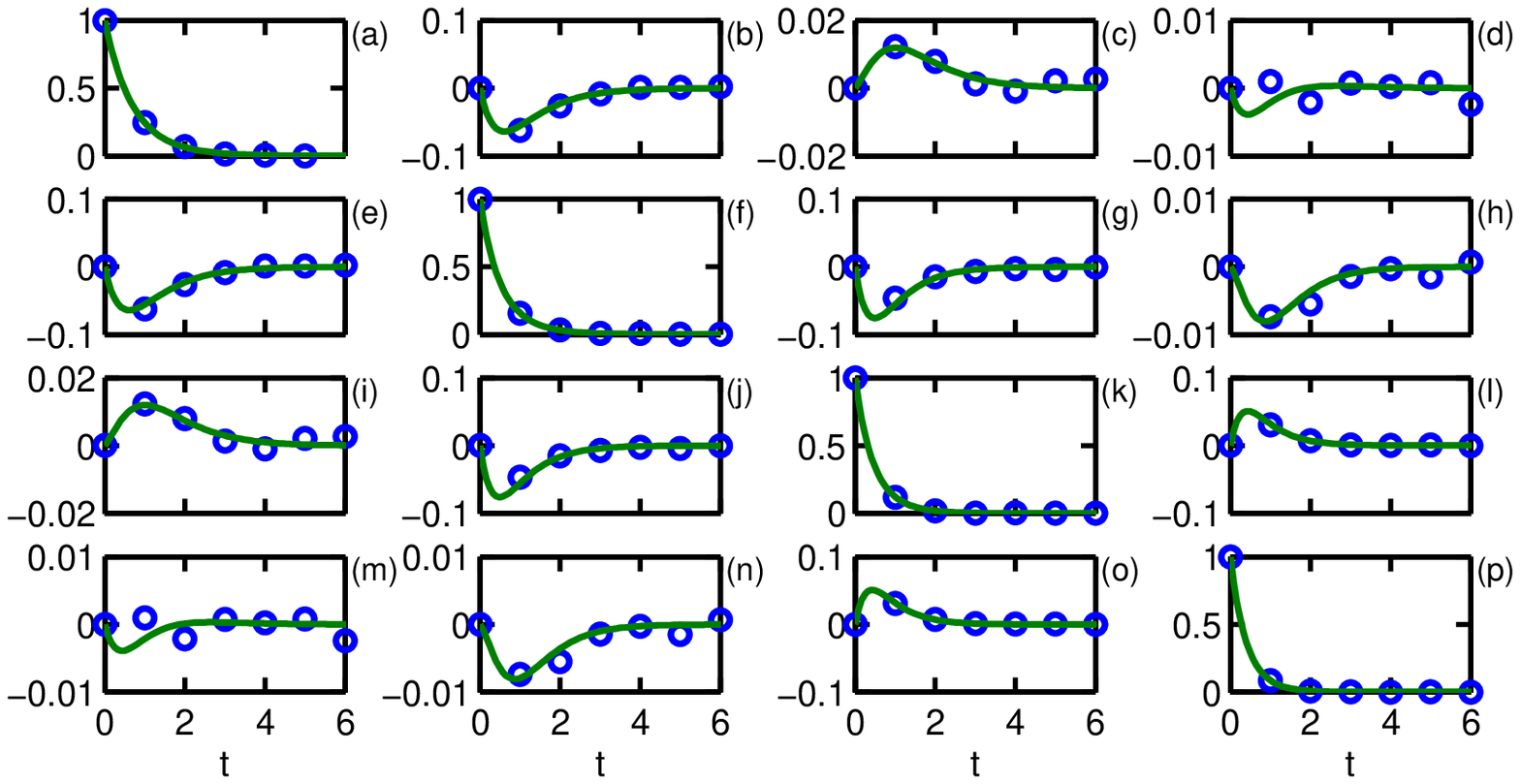}%
\caption{\label{fig:G} Comparison of $\tens{Y}_t \equiv Y_{ij}(t)$ for the model (solid) and data (circled) for the $0^+$ glueball operators using 10k data }
\end{figure*}

The best fitting parameters for our 10,000 measurement (10k) data are displayed in Table~\ref{tab:F} (with the standard error below the parameter value) for both torelon and $0^+$ glueball operators.  We have restricted these $\tens{Y}_t$ to $d=4$ to extract the lightest states, and the entire range of $t$ is included in the fit.  We also show the parameters for an independent run of 1,000 measurements (1k) using a similar but updated code.  (The 1k run averaged correlators over the $D=3$ independent temporal orientation thus is more like 2,500 measurements of the 10k run which did not.)  With only one model per set of data, there is no opportunity for model selection, but we include the value of the negative log evidence in the table for completeness.

The parameters of our model must be manipulated to produce estimates for the quantities in which we are interested, namely the masses and amplitudes of the energy eigenstates.  As the mass $m_j$ is the sum of the energy gaps $E_{i \leq j}$, its variance is given by the 1-norm of their covariance, $\sigma^2_m = \Vert \tens{\Sigma}_E \Vert_1$.  As there must be a high degree of correlation in $\tens{A}$ to enforce its orthogonality, and as the formula for $\vec{\Theta} \rightarrow \tens{A}$ is rather complicated, we have not computed standard errors for the entries of $\tens{A}$.  In Table~\ref{tab:G} we give the mass and amplitude parameters for our 10k and 1k runs.  These amplitudes, one recalls, are the overlaps with the orthogonalized $\tens{Y}_t$ data which themselves are a linear combination of the original basis $\tens{X}_t$.  We note the appearance of two negative real phases in $\tens{A}_\mathrm{box}$ between the 10k and 1k data for the states labelled 3 and 4 which explains why the final value of $\vec{\Theta}_\mathrm{box}$ differs in Table~\ref{tab:F}.

Finally, there is the graphical inspection of the quality of fit.  In Figs.~\ref{fig:F} and \ref{fig:G} we compare the $\tens{Y}_t$ of the model and the data for the torelon and $0^+$ glueball operators.  For either operator the model and data are in visibly close agreement, except for $Y_{14}^\mathrm{box}(t)$ which appears to be mostly noise.  The significant variation in the model lies primarily within the first quarter lattice extent, which here is spanned by only 5 data values---we simply are not working at sufficient temporal resolution to measure the cross-correlations where they vary most.  We also see that (for $m L \gg 1$) the relevant part of the data for fitting the model parameters lies well away from the midpoint of the lattice thus should not be much affected by the difference between $\exp$ and $\cosh$.  As a rule of thumb, if one's results depend upon the use of $\cosh$, then one is not working with a large enough lattice.


\begin{table}
\caption{Best fit parameters using 1k data for $0^+$ glueball operators including the vacuum contribution with the standard error below the parameter value}
\label{tab:H} 
\begin{tabular}{c|d{3}d{3}d{3}d{3}d{3}}
\hline\noalign{\smallskip}
$j$ & \multicolumn{1}{c}{0} & \multicolumn{1}{c}{1} & \multicolumn{1}{c}{2} & \multicolumn{1}{c}{3} & \multicolumn{1}{c}{4} \\
\noalign{\smallskip}\hline\noalign{\smallskip}
$E_j$ &   0.000 &  1.271 &  0.582 &  0.593 &  0.426 \\
$\sigma$ & 0.000 &  0.008 &  0.023 &  0.045 &  0.075 \\
\noalign{\smallskip}\hline\noalign{\smallskip}
$l$ & \multicolumn{1}{c}{1} & \multicolumn{1}{c}{2} & \multicolumn{1}{c}{3} & \multicolumn{1}{c}{4} & \multicolumn{1}{c}{5} \\
\noalign{\smallskip}\hline\noalign{\smallskip}
$\Theta_l$ &  -0.036 &  0.029 & -0.022 & -0.015 &  0.631 \\
$\sigma$ & 0.000 &  0.000 &  0.000 &  0.000 &  0.015 \\
\noalign{\smallskip}\hline\noalign{\smallskip}
$l$ & \multicolumn{1}{c}{6} & \multicolumn{1}{c}{7} & \multicolumn{1}{c}{8} & \multicolumn{1}{c}{9} & \multicolumn{1}{c}{10} \\
\noalign{\smallskip}\hline\noalign{\smallskip}
$\Theta_l$ &  -0.293 & -0.147 &  0.855 &  0.416 & -0.572 \\
$\sigma$ & 0.013 &  0.010 &  0.037 &  0.028 &  0.087 \\
\noalign{\smallskip}\hline
\end{tabular}
\end{table}

\begin{table}
\caption{Mass and amplitude parameters using 1k data for $0^+$ glueball operators including the vacuum contribution}
\label{tab:I} 
\begin{tabular}{c|d{3}d{3}d{3}d{3}d{3}}
\hline\noalign{\smallskip}
$j$ &\multicolumn{1}{c}{0} & \multicolumn{1}{c}{1} & \multicolumn{1}{c}{2} & \multicolumn{1}{c}{3} & \multicolumn{1}{c}{4} \\
\noalign{\smallskip}\hline\noalign{\smallskip}
$m_j$ &  0.000 &  1.271 &  1.852 &  2.446 &  2.872 \\
$\sigma$ &  0.000 &  0.008 &  0.022 &  0.041 &  0.068 \\
\noalign{\smallskip}\hline\noalign{\smallskip}
\multirow{5}{*}{$\tens{A}$}
&  0.999 & -0.052 &  0.010 & -0.005 & -0.001 \\
&  0.036 &  0.763 &  0.562 &  0.313 & -0.054 \\
& -0.029 & -0.559 &  0.332 &  0.735 & -0.190 \\
&  0.022 &  0.286 & -0.644 &  0.349 & -0.618 \\
&  0.015 &  0.147 & -0.400 &  0.489 &  0.761 \\
 \noalign{\smallskip}\hline
\end{tabular}
\end{table}

\subsection{Fitting the vacuum}
Our final application of the composite parametrization of the orthogonalized cross-correlators addresses whether one needs to work with vacuum subtracted operators for the $0^+$ glueball or not.  We note that the $\tens{X}_t$ need not be normalized, as Eqn.~(\ref{eqn:varidata}) produces a normalized $\tens{Y}_t$ by construction.  Neither must one take its vacuum subtracted value, as the vacuum is simply the ground state of operators with trivial quantum numbers.  To fit the vacuum contribution to $\tens{Y}_t$ without vacuum subtraction, we simply prepend the first gap parameter with one whose prior is uniform over $\pm 0.1$ and treat the first nonzero energy state as an excitation.  With this analysis we get for the 1k run of data the best fitting parameters shown in Table~\ref{tab:H}.  The vacuum energy evaluated to $E_0 = 0.32(35) \times 10^{-5}$, and this fit's quality is given by $\del_{10} L_E = -7.4$ and $-L_E = 209$.  From Table~\ref{tab:I} we see that the mass estimates for states above the vacuum are consistent with those in Table~\ref{tab:G} using vacuum subtracted correlators.

\section{Summary and conclusions}
\label{sec:concl}
\begin{figure}
\includegraphics{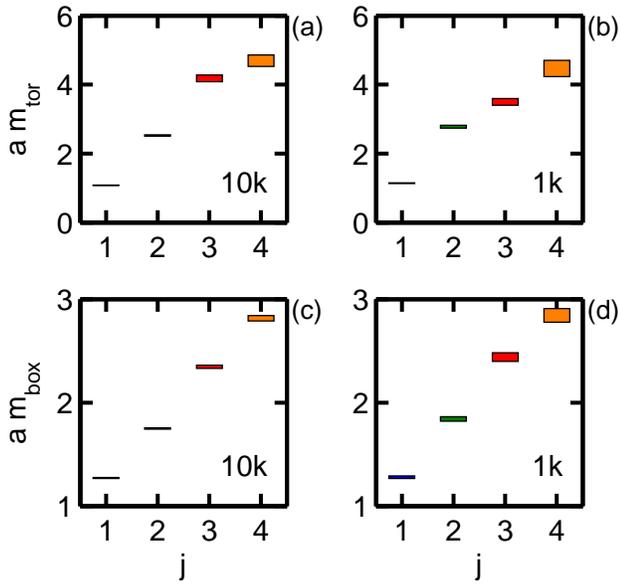}%
\caption{\label{fig:H} Mass spectra for torelon (a)-(b) and $0^+$ glueball operators (c)-(d) using the composite parametrization}
\end{figure}

We summarize our results for the mass eigenstates of the torelon and $0^+$ glueball on a $16^3$ lattice at $\beta=6$ using the composite parametrization of the orthogonalized cross-correlators in Fig.~\ref{fig:H}.  While there is a little bit of shifting of the estimates for the excited torelon states between the 10k and 1k data, the estimates for the $0^+$ glueball are consistent for the long and short measurement runs.  The expectation of a discrete spectrum is observed for these lightest few energy eigenstates.  While there is an improvement in the error bar width for the 10k data, the estimate from the 1k data is already fairly precise.  (One of course should not confuse precision with accuracy.)  By obviating the need to normalize and vacuum subtract one's timeslice correlators, there should be a modest decrease in one's execution time (not investigated).  While ``Bayesian methods are not a cure for bad data''~\citep{Morningstar:2002185}, they do let one get the most out of the data which one has on hand.

Fitting a sum of exponentials to a noisy correlation function is a hard problem.  Without the recognition of the orthogonal contributions, one is left with the methodology of Section~\ref{sec:torelon}, which was only successful for the simplest of cases.  Nonetheless, the use of a non-uniform prior offers the best chance to turn an intractable problem into one that is soluble.  With the composite parametrization used in Section~\ref{sec:compofit}, we have successfully fit mass and amplitude parameters to the orthogonalized cross-correlation functions produced by a lattice gauge theory simulation.  The models presented here are well defined, and their use by others is encouraged.  We are curious how well they might perform given better quality simulation data with greater temporal resolution.

In conclusion, we have considered various models of exponential decay applicable to lattice correlation functions.  The evaluation of the merit function includes contributions from the non-uniform priors appropriate for the amplitude and mass as well as the measure of fit.  Analysis of torelon and glueball correlation functions indicates that a sum of exponentials is present in the data even after application of an orthogonalizing procedure.  The use of maximal evidence rather than maximal likelihood parameter estimation is encouraged for the extraction of excited states from lattice correlation functions.


\section*{Acknowledgments}
The author appreciates occasional conversations with Mike Teper on the use of lattice gauge theory and with Christoph Spengler on the use of the composite parametrization.

\bibliographystyle{unsrt}

\end{document}